\documentclass[prb,twocolumn,showpacs,preprintnumbers,amsmath,amssymb,superscriptaddress,]{revtex4}
\usepackage{graphics}
\usepackage{graphicx}
\usepackage{dcolumn}
\usepackage{bm}
\usepackage{amsthm}
\usepackage{amsmath}
\usepackage{amsfonts}
\usepackage[dvips]{color}

\begin{document}


\title{Strong-coupling solution of the bosonic dynamical mean-field theory} 

\author{Anna Kauch}
\email{kauch@fzu.cz}

\affiliation{Institute of Physics, Academy of Sciences of the Czech Republic, Na Slovance 2, 18221 Praha, Czech Republic}

\author{Krzysztof Byczuk}%
 \affiliation{%
Institute of Theoretical Physics, Faculty of Physics, University of Warsaw, Ho\.za 69, PL-00-681 Warszawa, Poland }

\author{Dieter Vollhardt}
 \affiliation{Theoretical Physics III, Center for Electronic Correlations and Magnetism, Institute of Physics, University of Augsburg, D-86135 Augsburg, Germany}

\date{\today}

\begin{abstract}

We derive an approximate analytical solution of the self-consistency equations of the bosonic dynamical mean-field theory (B-DMFT) in the strong-coupling limit. The approach is based on a linked-cluster expansion in the hybridization function of normal bosons around the atomic limit. The solution is used to compute the phase diagram of the bosonic Hubbard model for different lattices. We compare our results with numerical solutions of the B-DMFT equations and numerically exact methods, respectively. The very good agreement with those numerical results demonstrates that our approach captures the essential physics of correlated bosons both in the Mott insulator and in the superfluid phase. Close to the transition into the superfluid phase the momentum distribution function  at zero momentum is found to be strongly enhanced already in the normal phase. The linked-cluster expansion also allows us to compute dynamical properties such as the spectral function of bosons. The evolution of the spectral function across the transition from the normal to the superfluid phase is seen to be characteristically different for the interaction driven and density driven transition, respectively.

\end{abstract}


\maketitle

\section{Introduction}

Cold atoms in optical lattices provide a fascinating new class of interacting quantum many-particle systems.\cite{bec,bloch_rmp} Due to the unprecedented precision of experimental techniques in this field it is now possible to simulate and experimentally test theoretical models.\cite{Greiner02,optical_bmit,lewenstein07,optical_mit,giorgini_rmp,bloch_rmp} In particular, experiments with bosonic atoms have revived the theoretical interest in the properties of the bosonic Hubbard model.\cite{fisher,jaksch,refs} This model describes the quantum mechanical competition between the kinetic energy of lattice bosons, which is responsible for their Bose-Einstein condensation, and the repulsive interaction, which favors localization of the particles. The phase diagram of the bosonic Hubbard model was first
calculated by Fisher \emph{et al.}~\cite{fisher} within a static mean-field theory derived from the atomic limit.
With the formulation of the bosonic dynamical mean-field theory (B-DMFT)~\cite{bdmft,bose_fermi} a~comprehensive investigation scheme for correlated lattice bosons in the thermodynamic limit has become available, which allows one to calculate also dynamical properties such as spectral functions of the interacting bosons. The B-DMFT is a thermodynamically consistent, non-perturbative many-body approach which is applicable for all values of the input parameters, e.g., the interaction, density, and temperature. It leads to a set of nonlinear equations which need to be solved self-consistently. An exact solution can be found only in special cases, e.g., for the Falicov-Kimball model.\cite{bdmft,bose_fermi} In general, the self-consistent equations have to be solved numerically or by employing approximate analytical methods. The experience with the fermionic DMFT~\cite{metzner89,dmft_phys_today,dmft_review} shows that both numerically exact (but computationally expensive) methods \emph{and} approximate analytical methods are important to gain insight into the solution of the complicated self-consistency equations. So far solutions of the B-DMFT equations had to be obtained fully numerically. Hu and Tong,\cite{ninghua} and Hubener, Snoek, and Hofstetter\cite{hofstetter} employed exact diagonalization (ED), and Anders \emph{et al.}\cite{werner,werner2} made use of continuous-time quantum Monte Carlo (CT-QMC) to solve the B-DMFT equations. Analytical or semi-analytical solutions of the B-DMFT equations did not exist up to now.

In this paper we present an analytical strong-coupling solution of the B-DMFT derived by a linked-cluster expansion (LCE)\cite{metzner91,bartkowiak} around the atomic limit. The method is analogous to the fermionic strong-coupling solver developed by Dai, Haule, and
Kotliar\cite{kotliar_sc} for the fermionic DMFT. While in the fermionic case the strong-coupling expansion is unable to capture the low temperature Fermi liquid physics due to the existence of a characteristic low energy (Kondo) scale, there is no such limitation in the bosonic case. Our approach differs from previous strong-coupling expansions to the bosonic Hubbard model\cite{kampf,sengupta,pelster1,pelster,freericks} since they performed the expansion in the hopping amplitude.

The paper is organized as follows. We first introduce the B-DMFT and its self-consistency equations. Then we formulate the linked-cluster expansion and thereby derive  a strong-coupling approximation to the B-DMFT equations. This is then applied to the Bethe lattice and the cubic lattice, both with coordination number $z=6$, and to the Bethe lattice with $z=\infty$. The phase diagrams of the bosonic Hubbard, model calculated in this way are compared with those obtained from numerical solutions of the B-DMFT computed with ED \cite{hofstetter} and CT-QMC,\cite{werner,werner2} respectively, from numerically exact evaluations on a Bethe lattice,\cite{zamponi} and from numerical results obtained by direct Monte Carlo simulations of the bosonic Hubbard model.\cite{qmc} The momentum distribution functions
and spectral functions  of correlated lattice bosons in the normal and the Bose-Einstein condensed phase are also calculated. Finally we discuss possible extensions  of the approach.

\section{Cumulant expansion in the bosonic dynamical mean-field theory}

The B-DMFT is the bosonic counterpart to the well-established DMFT for lattice fermions described by the Hubbard model. Its derivation is described in detail in Ref.~\onlinecite{bdmft}. Here we focus on a single species of bosons. The expansion presented below is easily generalized to the case of more than one type of boson.

The bosonic Hubbard model is given by the Hamiltonian
\begin{equation}
H=\sum_{ij}t_{ij} b^{\dagger}_{i} b_{j}+\frac{1}{2}U \sum_{i} n_{i}( n_{i}-1),
\label{BH_model}
\end{equation}
where $b_{i}^{\dagger }$ and $b_{i}$ are creation and annihilation
operators, respectively, for a boson at a lattice site $i$, $t_{ij}$ is the hopping between lattice sites $i$ and $j$, $U$ is the local interaction, and $n_{i}=b^{\dagger}_{i} b_{i}$ is the number operator of the local occupation.  In this paper we consider nearest-neighbor hopping, i.e., $t_{ij}=-t$ for the nearest-neighbor sites $i$,$j$, and $t_{ij}=0$ otherwise.
In the following we set the Boltzmann constant $k_B$ and the lattice spacing $a$ equal to unity.

\subsection{Local action of the B-DMFT}

In the B-DMFT the $d$-dimensional lattice problem \eqref{BH_model} is replaced by an effective single-site (``impurity'') problem in which the local interaction $U$ remains unchanged, but the rest of the lattice is replaced by two dynamical mean fields (``baths'') corresponding to bosons in the normal state and in the Bose-Einstein condensate, respectively.\cite{bdmft}
The time evolution of bosons on a particular site
$i=0$ is represented by the local Green function
\begin{equation}
\mathbf{G}(\tau)=- \langle T_{\tau} \mathbf{b} (\tau)
\mathbf{b}^{\dagger}(0) \rangle_{S_{\mathrm{loc}}},
\label{G_local}
\end{equation}
where we used the imaginary time, finite temperature formalism and Nambu
notation with
\begin{equation}
{\bf b} = \left(
\begin{array}{l}
b\\
b^{*}
\end{array}
\right),
\end{equation}
and the Bose-Einstein condensate (BEC) is described by the local order parameter
 \begin{equation}
\phi=\langle  {b}(\tau) \rangle_{S_{\mathrm{loc}}}.
\label{phi_local}
\end{equation}
The impurity problem is defined by the local action
\begin{eqnarray}
S_{\mathrm{loc}}= \int_0^{\beta} \!\!\!\!d \tau  b^{*} (\tau) (\frac{\partial}{\partial \tau} - \mu) b (\tau) + \frac{1}{2}\int_0^{\beta}\!\!\!\! d \tau U n(\tau)(n(\tau)-1)  \nonumber \\ + \kappa\int_0^{\beta} \!\!\!\!d \tau  {\pmb \Phi}^{\dagger}(\tau) {\bf b} (\tau)
+\frac{1}{2}\int_0^{\beta}\!\!\!\! d \tau  \int_0^{\beta} \!\!\!\! d \tau'  \,\,{\bf b}^{\dagger} (\tau){\pmb \Delta}(\tau-\tau')  {\bf b}(\tau'),
\label{S_dmft}
\end{eqnarray}
where $\mu$ is the chemical potential, $\kappa=\sum_{i\neq 0} t_{i0}$ is a lattice dependent parameter, and
\begin{equation}
\quad {\pmb \Phi } = \left(
\begin{array}{l}
\Phi \\
\Phi^*
\end{array}
\right)
\end{equation}
is the condensate wave function, i.e., a dynamical mean field. The dynamical mean field corresponding to bosons in the normal state is represented by the hybridization function
\begin{equation}{\pmb \Delta }(\tau-\tau')= \left(
\begin{array}{cc}
\Delta_{11}(\tau-\tau') & \Delta_{12}(\tau-\tau') \\
\Delta_{21}(\tau-\tau') & \Delta_{22}(\tau-\tau')
\end{array}
\right).
\end{equation}
The dynamical mean fields $\Phi(\tau)$, $\Delta_{11}(\tau)$, and $\Delta_{12}(\tau)$ are determined by the self-consistency equations
\begin{eqnarray}
\pmb{\Delta}(\tau-\tau')&=&-\sum_{i,j\neq0} t_{i0}t_{0j}\langle T_{\tau}{\bf b}_i(\tau ) {\bf b}_j^{\dagger}(\tau')\rangle_{(0)}\nonumber \\ &=&\sum_{i,j\neq0} t_{i0}t_{0j} {\bf G}_{ij}^{(0)}(\tau-\tau')
\label{sc_delta}
\end{eqnarray} and
\begin{equation}
\pmb{ \Phi}=\langle {\bf b}(\tau )\rangle_{(0)}.
\label{sc_phi}
\end{equation}
Here the notation $\langle \cdots \rangle_{(0)}$ indicates that the thermodynamic average is performed on a lattice with a cavity, i.e., with one site removed.
We note that in equilibrium $\pmb{ \Phi}(\tau)$ is constant.
For finite dimensional lattices $\pmb{ \Phi}$ is related to the local BEC order parameter \eqref{phi_local} by
\begin{equation}
\pmb{\Phi}=\left(\!
\begin{array}{cc}
\!1\!-\!\frac{1}{\kappa}\! \int\!\!d \tau \Delta_{11}(\tau) & \!-\frac{1}{\kappa}\!\int \!\!d \tau\Delta_{12}(\tau) \smallskip\\
\!-\frac{1}{\kappa}\!\int \!\!d \tau\Delta_{21}(\tau) & \!1\!-\!\frac{1}{\kappa}\!\int \!\!d \tau\Delta_{22}(\tau)
\end{array}
\!\right)\!\!\left(\!
\begin{array}{l}
\phi \!\\ \phi^*\!
\end{array}\!\right).
\label{sc_phi_2}
\end{equation}

The self-consistency loop is closed by introducing the self-energy in the Matsubara frequency representation through the ${\bf k}$-integrated Dyson equation
\begin{equation}
{\pmb{\Sigma}(i\omega_n)}=
 \left(
\begin{array}{cc}
i \omega_n \!+\!\mu & 0 \\
0 & -i\omega_n \!+\! \mu
\end{array}
\right)
- {\pmb{\Delta}(i\omega_n)}-[{\bf G}(i\omega_n)]^{-1}
\label{sc_dyson}
\end{equation}
and using the lattice Hilbert transform
\begin{equation}
{\bf G}(\!i\omega_n\!)\!=\!\! \int \!\!N_0(\epsilon) \!
\left[\!\left(\!
\begin{array}{cc}
\!i\omega_n\! +\! \mu\! -\! \epsilon \! & 0 \\
0 & \!-i\omega_n \! +\!\mu\! -\! \epsilon\!
\end{array}
\!\right)
\!-\!{\pmb{\Sigma}(\!i\omega_n\!) }\!\right]^{-1}.
\label{sc_hilbert}
\end{equation}
The latter equation links the local Green function to the self-energy for a specific lattice described by the non-interacting density of states $N_0(\epsilon)$. The momentum dependent lattice Green function ${\bf G}({\bf k},i\omega_n)$ is then given by
\begin{equation}
{\bf G}({\bf k},i\omega_n)\!=\!
\left[\!\left(\!
\begin{array}{cc}
\!i\omega_n \!+\! \mu\! -\! \epsilon_{\bf k}\!  & 0 \\
0 &\! -\!i\omega_n\!+\!\mu\! -\! \epsilon_{\bf k}\!
\end{array}
\!\right)
\!-\!{\pmb{\Sigma}(\!i\omega_n\!) }\!\right]^{-1},
\label{g_k}
\end{equation}
where $\epsilon_{\bf k}$ is the dispersion relation of the non-interacting system and ${\pmb{\Sigma}}(i\omega_n) $ is the self-consistent solution of equations \eqref{G_local}-\eqref{sc_hilbert}.

For a Bethe lattice with infinite connectivity ($z=\infty$) \cite{bethe_lattice2, bethe_lattice1} the self-consistency conditions reduce to the simple expressions $\pmb{\Delta}(\tau-\tau')=t^2{\bf G}(\tau-\tau')$
and  $\pmb{ \Phi} = (\phi,\phi^*).$ In general, e.g., for a cubic lattice, the self-consistency equations \eqref{sc_phi_2}-\eqref{sc_hilbert} need to be solved numerically.

\subsection{Cumulant expansion}

In order to solve the impurity problem defined above we use the cumulant (linked-cluster) expansion in the dynamical mean fields $\Delta_{11}$ and  $\Delta_{12}$. The action \eqref{S_dmft} is further divided into two parts
\begin{equation}
S_\mathrm{loc}=S_0+S',
\end{equation}
where
\begin{eqnarray}
S_0&=&\int_0^{\beta} \!\!\!\!d \tau  b^{*} (\tau) (\frac{\partial}{\partial \tau} - \mu) b (\tau) + \frac{1}{2}\int_0^{\beta} d \tau U n(\tau)(n(\tau)-1)  \nonumber \\ &+&\kappa\int_0^{\beta} \!\!\!\!d \tau  { \Phi}^{\dagger}(\tau) {\bf b} (\tau)
\end{eqnarray}
and
\begin{equation}
S'=\frac{1}{2}\int_0^{\beta}\!\!\!\! d \tau  \int_0^{\beta} \!\!\!\! d \tau'  \,\,{\bf b}^{\dagger} (\tau){\Delta}(\tau,\tau')  {\bf b}(\tau').
\end{equation}
The partition function of the impurity problem $Z$ is thus written as
\begin{equation}
Z = Z_{0}\langle e^{- S'} \rangle_{0},
\label{Z_imp}
\end{equation}
where $Z_{0}$ is the partition function for the system described by $S_0$, and $\langle ...\rangle_{0}$ denotes the thermodynamic average with respect to the action $S_0$.

Now the exponential function appearing in the average is expanded, leading to an infinite series 
\begin{eqnarray}
\langle e^{- S'}\rangle_{0} \!\!&\!=\!&\!1\!-\!\frac{1}{2}\int_0^{\beta}\!\!\!\! d \tau \!\!\!\int_0^{\beta} \!\!\!\! d \tau'  \,\,\langle T_{\tau}{\bf b}^{\dagger} (\tau){\Delta}(\tau,\tau')  {\bf b}(\tau') \rangle_{0} \nonumber \\
&\!+\!&\!\frac{1}{4\cdot2!}\int_0^{\beta}\!\!\!\! d \tau_1 \!\!\! \int_0^{\beta} \!\!\!\! d \tau'_1\!\!\! \int_0^{\beta} \!\!\!\! d \tau_2  \!\!\!\int_0^{\beta}\!\!\!\!  d \tau'_2  \,\,\langle T_{\tau}{\bf b}^{\dagger} (\tau_1){\Delta}(\tau_1,\tau'_1) \nonumber\\ &\times & {\bf b}(\tau'_1)\,\, {\bf b}^{\dagger} (\tau_2){\Delta}(\tau_2,\tau'_2) {\bf b} (\tau'_2)\rangle_{0}  + \ldots
\end{eqnarray}
The series is then re-exponentiated with the help of cumulants (i.e., connected $n$-particle Green functions)
 \cite{kubo,metzner91}
\begin{eqnarray}
\langle e^{- S'}\rangle_{0}\!\! &\!=\!&\!\exp \left\{-\frac{1}{2}\int_0^{\beta}\!\!\!\! d \tau\!\!\! \int_0^{\beta} \!\!\!\! d \tau'  \,\,\langle T_{\tau}{\bf b}^{\dagger} (\tau){\Delta}(\tau,\tau')  {\bf b}(\tau') \rangle_{0}^{c} \right. \nonumber \\
&+&\frac{1}{4\cdot2!}\int_0^{\beta}\!\!\!\! d \tau_1\!\!\!  \int_0^{\beta} \!\!\!\! d \tau'_1\!\!\! \int_0^{\beta} \!\!\!\! d \tau_2\!\!\!  \int_0^{\beta}\!\!\!\!  d \tau'_2 \,\,\langle T_{\tau}{\bf b}^{\dagger} (\tau_1){\Delta}(\tau_1,\tau'_1) \nonumber \\
 &\!\times\! &\! \left. {\bf b} (\tau'_1)\,\, {\bf b}^{\dagger} (\tau_2){\Delta}(\tau_2,\tau'_2) {\bf b} (\tau'_2)\rangle_{0}^{c}  + \ldots\right\}.
\end{eqnarray}
Here the superscript $c$ indicates that only the connected part of the averages with respect to $S_0$ is included. Now the partition function \eqref{Z_imp} can be calculated to the desired order in $\Delta$.

The above approximation is in the spirit of other strong-coupling expansions \cite{freericks,pelster} and becomes exact in the atomic limit ($\Delta_{11}=\Delta_{12}=0$, $\phi=0$). However, it should be stressed that it is \emph{not} an expansion in the hopping amplitude but rather in the dynamical mean fields $\Delta_{11}$ and  $\Delta_{12}$. The fact that these fields are obtained self-consistently implies that all orders of the hopping amplitude contribute.\cite{bdmft,metzner91}

In the following we perform the cumulant expansion to second order in $\Delta_{11}$ and  $\Delta_{12}$ in the partition function $Z$. Since the Green function is determined by the functional derivative
\begin{equation}
G_{\alpha \beta}(\tau-\tau')=-2\frac{\delta \ln Z }{\delta \Delta_{\alpha \beta}(\tau',\tau)},
\end{equation}
the diagonal  element $G_{11}(\tau-\tau')$ and off-diagonal element $G_{12}(\tau-\tau')$
are then of first order in $\Delta_{11}$ and  $\Delta_{12}$:
\begin{eqnarray}
G_{11}(\tau-\tau')= -\langle T_{\tau}  b(\tau )b^{*}(\tau') \rangle_{0} \quad \quad  \quad \nonumber \\
 +\frac{1}{2}\int_0^{\beta}\!\!\!\!d \tau_1  \!\!\! \int_0^{\beta} \!\!\!\! d \tau'_1   \,\, \langle T_{\tau}  b(\tau ){\bf b}^{\dagger} (\tau_1){\pmb \Delta}(\tau_1,\tau'_1)  {\bf b}(\tau'_1) b^{*}(\tau' )\rangle_{0}^{c},
\end{eqnarray}
and
\begin{eqnarray}
G_{12}(\tau-\tau')= -\langle T_{\tau}  b(\tau )b(\tau') \rangle_{0}\quad \quad  \quad  \nonumber \\
+ \frac{1}{2}\int_0^{\beta}\!\!\!\!d \tau_1 \!\!\!  \int_0^{\beta} \!\!\!\! d \tau'_1 \,\,\langle T_{\tau} b(\tau ){\bf b}^{\dagger} (\tau_1){\pmb \Delta}(\tau_1,\tau'_1)  {\bf b}(\tau'_1) b(\tau' )\rangle_{0}^{c}.
\end{eqnarray}
Furthermore, the local BEC order parameter is given by
\begin{equation}
\phi= \langle b(\tau) \rangle_{0}+\frac{1}{2}\int_0^{\beta}\!\!\!\!d \tau_1  \!\!\! \int_0^{\beta} \!\!\!\! d \tau'_1   \,\,\langle T_{\tau} b(\tau ){\bf b}^{\dagger} (\tau_1){\pmb \Delta}(\tau_1,\tau'_1)  {\bf b}(\tau'_1) \rangle_{0}^{c}.
\end{equation}

The thermodynamic averages are performed as $\langle \cdots \rangle_{0}= \frac{1}{Z_0}Tr(e^{-\beta H_0} \cdots)$,
with $H_{0}=\frac{1}{2} U  n(n-1)-\mu n + \kappa(\Phi b^{\dagger} + \Phi^* b)$. The trace is calculated over the eigenstates of $H_0$, which are obtained by an exact diagonalization of the Hamiltonian matrix which is represented in the occupation number basis. Since the local Hilbert space of $H_0$ for the bosonic impurity problem is infinite dimensional, the diagonalization has to be performed numerically which, in principle, implies a further approximation. The Hilbert space has to be cut off in the occupation number of the impurity. The error introduced thereby can be controlled by performing calculations with different values of the cut-off and choosing the smallest cut-off value such that the results do not differ within the required accuracy.\cite{cm2}

\section{Application of the linked-cluster expansion to various lattices}

In the following we apply the results of the LCE to the Bethe lattice and the cubic lattice, both with coordination number $z=6$, as well as to the Bethe lattice with infinite connectivity ($z=\infty$). Our results for the Bethe lattice with coordination number $z< \infty$ can benchmarked by the exact numerical solution based on the cavity method.\cite{zamponi}

\begin{figure}
        \includegraphics[width=0.45\textwidth]{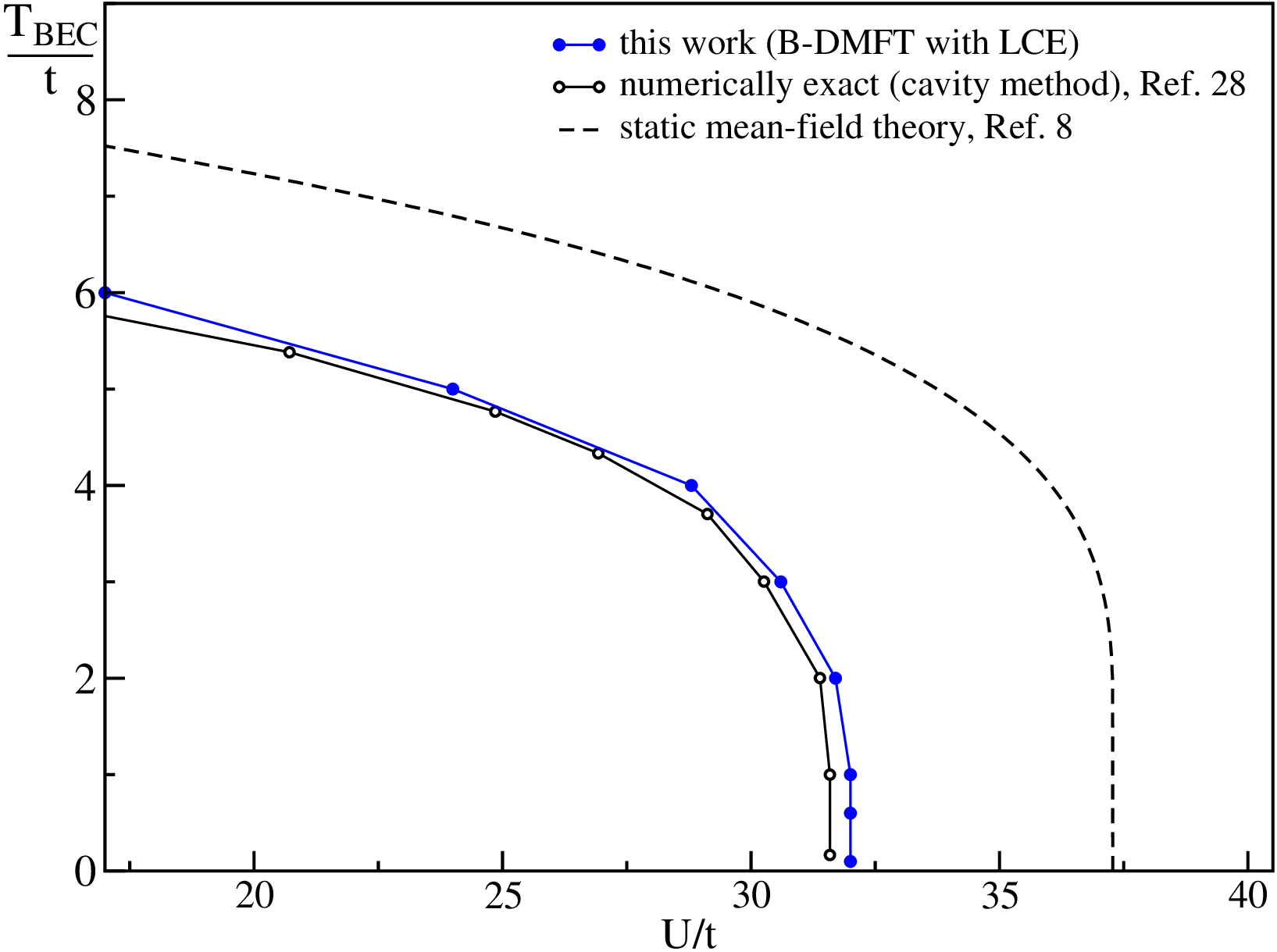}\hfill        \includegraphics[width=0.45\textwidth]{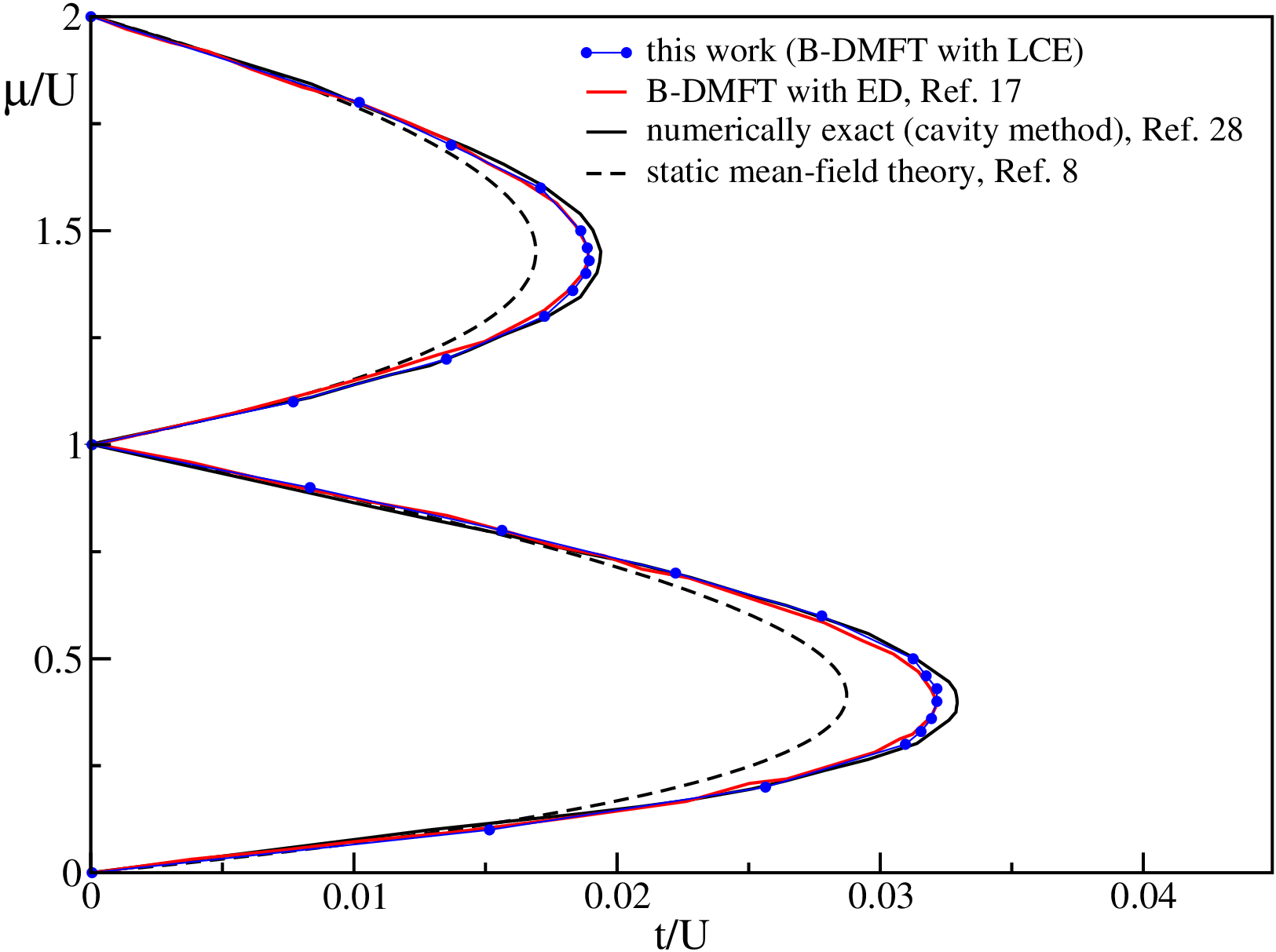}
\caption{ Results for the Bethe lattice with coordination number $z=6$. Top:  Dependence of the condensation temperature $T_\mathrm{BEC}$  on the interaction $U$ at density $\langle n \rangle = 1$. Bottom: Phase diagram $\mu/U$ \emph{vs.} $t/U$ computed by different methods: B-DMFT solved with LCE (this work, $T=0.1t$), B-DMFT solved with ED ($T=0$),\cite{hofstetter} numerically exact evaluation (cavity method)  ($T \le0.25t$),\cite{zamponi} and static mean-field solution ($T=0$).\cite{fisher} Inside the Mott lobes the system is Mott insulating and the occupation number is integer, while outside the system is superfluid.}
\label{bethe6}
\end{figure}

\subsection{Bethe lattice with coordination number $z=6$}

  In Fig.~\ref{bethe6} we show the results obtained with the LCE for the interaction dependence of the Bose-Einstein condensation temperature $T_\mathrm{BEC}(U)$, as well as for the phase diagram $\mu/U$ \emph{vs.} $t/U$ at $T=0.1t$. We also compare them with results from other methods: the exact numerical evaluation (cavity method) by Semerjian, Tarzia, and  Zamponi,\cite{zamponi} the B-DMFT solution with ED by Hubener, Snoek, and Hofstetter,\cite{hofstetter} and the static mean-field solution of Fisher \emph{et al.}~\cite{fisher} The static mean-field and the ED results were calculated at $T=0$, whereas the results of the cavity method were obtained for $T \le0.25t$. The phase transition line $\mu/U$ \emph{vs.} $t/U$ only weakly depends on $T$ at such low temperatures as can be seen in the upper panel of Fig.~\ref{bethe6}, where below $T_\mathrm{BEC}(U)/t=1$ the curve is practically vertical. For this reason we conclude that the phase diagram presented in the lower  panel of Fig.~\ref{bethe6} is essentially the ground state phase diagram.

The results shown in the lower  panel of  Fig.~\ref{bethe6} demonstrate that the agreement between the two B-DMFT solutions is excellent. Namely, the blue circles (LCE, this work) are seen to lie practically on the red line (ED from Ref.~\onlinecite{hofstetter}). Apparently the transition from the Mott-insulator to the superfluid is well described by the LCE approximation, which expands to first order in the dynamical mean field $\pmb{\Delta}(\tau)$. This is different from the case of the fermionic DMFT where the low temperature physics of the Hubbard model close to the metal-insulator transition can not be described by the strong-coupling approximation.\cite{kotliar_sc}

The value of the transition temperature $T_\mathrm{BEC}(U)$ obtained by the B-DMFT and the cavity method, respectively, is significantly lower than the results obtained by the static mean-field theory.\cite{fisher} Since the B-DMFT captures local fluctuations exactly we conclude that they are responsible for the lowering of $T_\mathrm{BEC}(U)$ and the associated increase of the size of the Mott lobes.

For strong interactions the system is a Mott insulator for most values of the chemical potential $\mu$. Upon lowering the interaction the system enters the superfluid phase with an order parameter $\phi\neq0$. For the values of the chemical potential between the Mott lobes the superfluid phase persists up to very large values of $U$. Since the LCE calculations were performed at a low but finite temperature ($T=0.1t$),  there is no superfluid phase at $\mu=U$ below $t/U\approx 0.0001$ (not discernible in the figure).

\subsection{Cubic lattice}

\subsubsection{Phase diagram}

\begin{figure}
\includegraphics[width=0.45\textwidth]{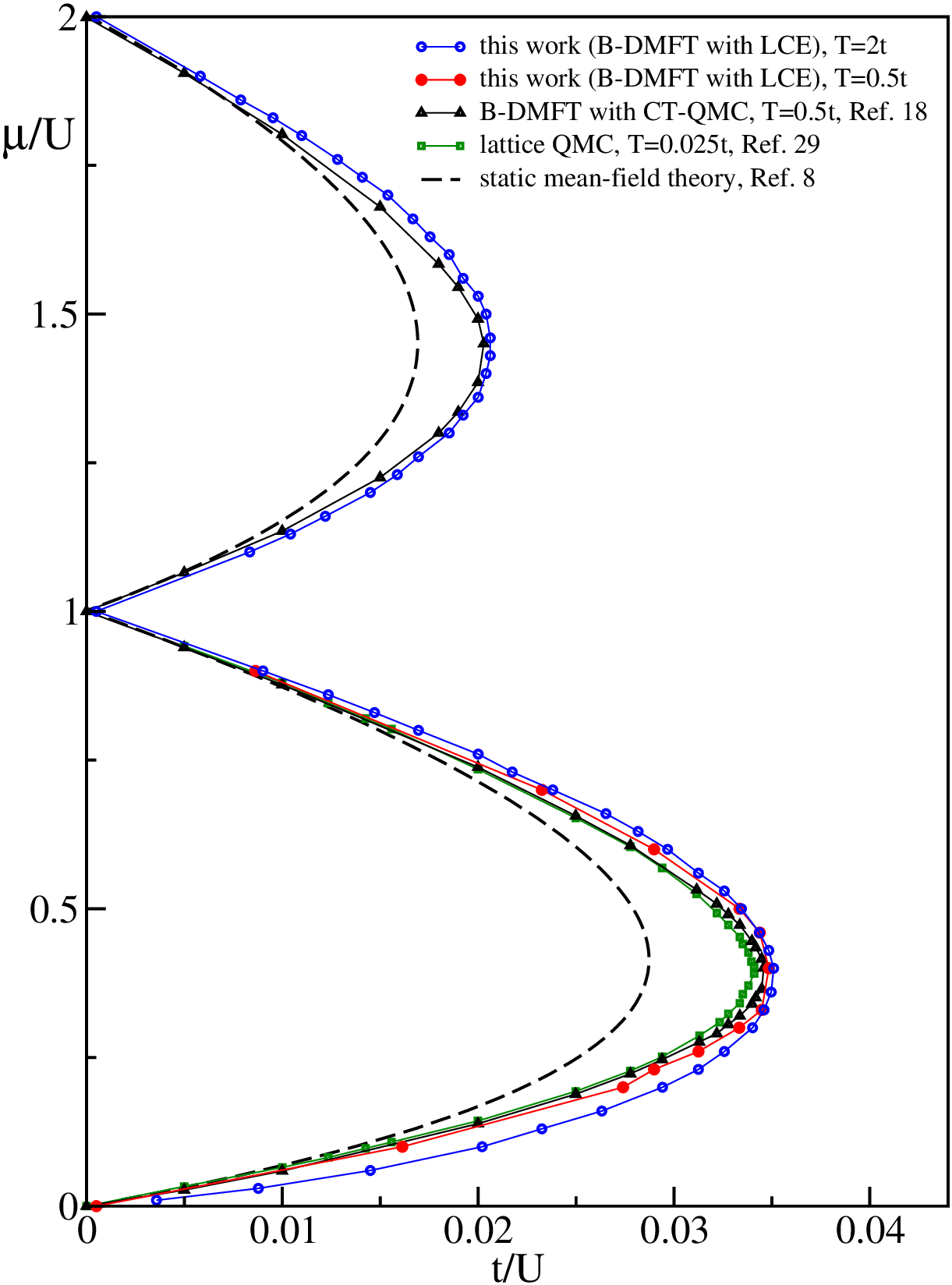}
\caption{Phase diagram $\mu/U$ \emph{vs.} $t/U$ for the cubic lattice obtained from B-DMFT with LCE compared with the results obtained from B-DMFT with CT-QMC (data from Ref.~\onlinecite{werner}), lattice QMC (data from Ref.~\onlinecite{qmc}), and static mean-field theory.\cite{fisher}}
\label{cubic}
\end{figure}

 The phase diagram of the Bose-Hubbard model for the cubic lattice obtained from the B-DMFT with the LCE and with CT-QMC, respectively, is presented in Fig.~\ref{cubic}. These results are compared with the lattice quantum Monte Carlo (QMC) results.\cite{qmc} The LCE results are shown for two different temperatures ($T=2t$ and $T=0.5t$). It is evident that the size of the Mott lobes decreases with decreasing temperature. Upon lowering the temperature the computation of the phase boundary using the B-DMFT with the LCE was found to become  more elaborate. As already noted in Ref.~\onlinecite{werner2} for the CT-QMC solver, the convergence of the DMFT cycle close to the phase transition is very slow  and the initial guess of $\pmb{\Delta}(\tau)$ and  $\pmb{\Phi}$ has to be carefully chosen.

Fig.~\ref{cubic} shows that there is a small quantitative difference between the results obtained by different methods. It is unlikely that these differences can be explained by the different temperatures used in the computations (the lattice QMC calculations~\cite{qmc} were performed at $T=0.025t$, which is lower than the temperature used in the B-DMFT calculations). Indeed, at such low temperatures the temperature dependence of the phase diagram is very weak, as discussed earlier for the $z=6$ Bethe lattice. Nevertheless, the overall agreement between the results obtained from the three different methods is clearly very good. As in the case of the $z=6$ Bethe lattice the local dynamical fluctuations described by the B-DMFT lead to an increase of the size of the Mott lobes compared to the static mean-field solution.

\begin{figure}
        \includegraphics[width=0.45\textwidth]{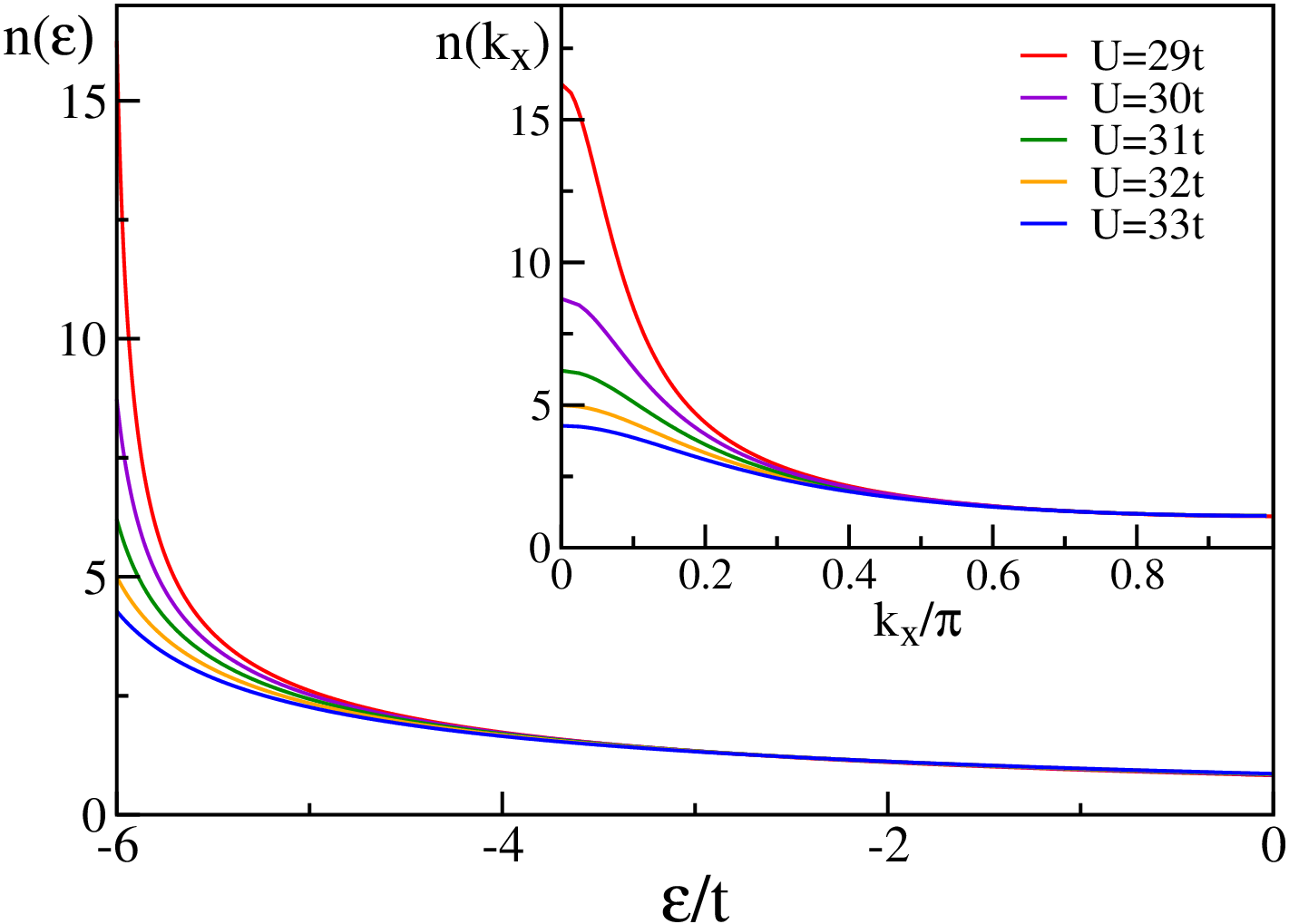} \hfill
        \includegraphics[width=0.45\textwidth]{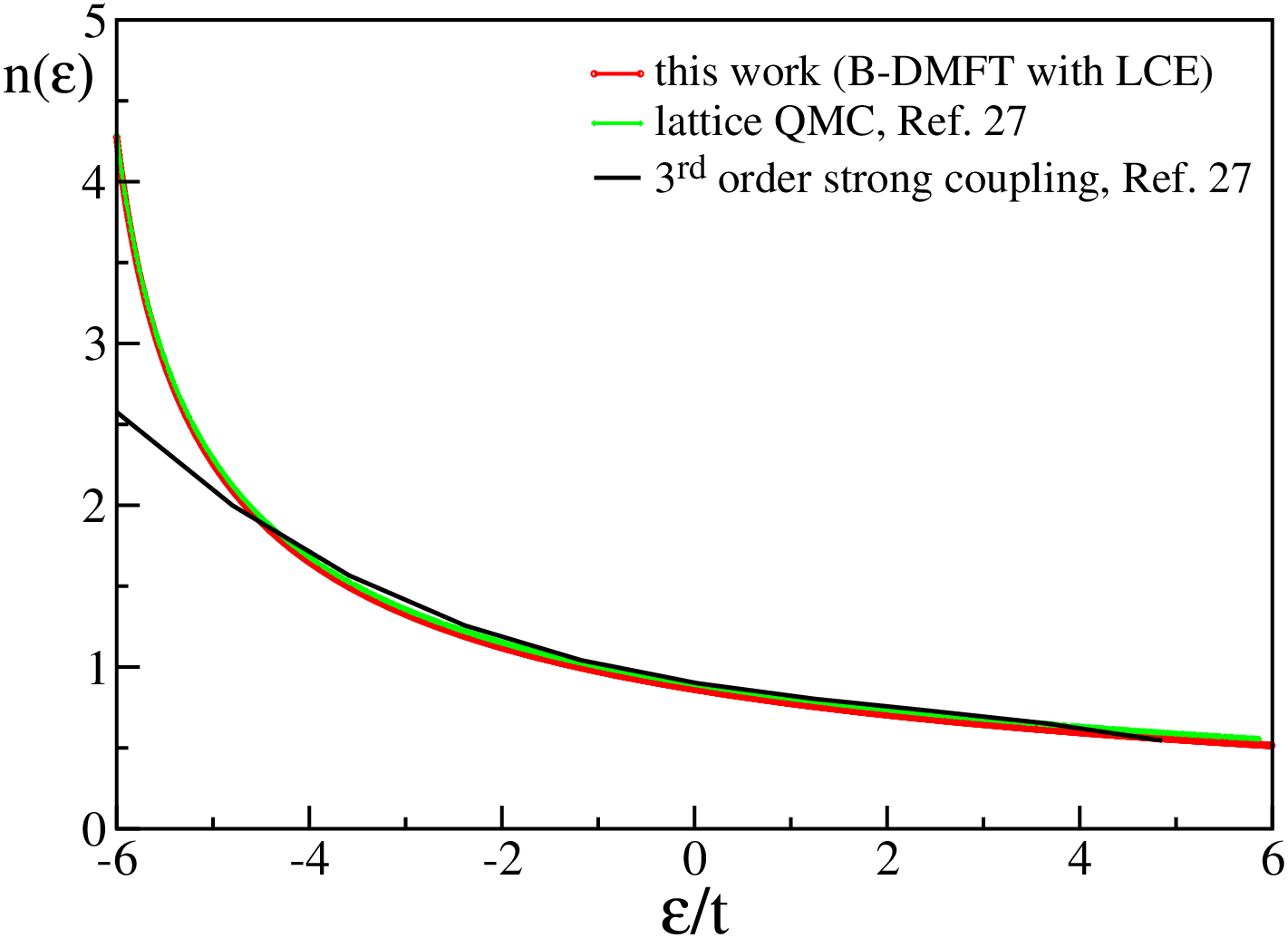}
\caption{Momentum distribution function for the cubic lattice in the normal phase calculated with the LCE at temperature $T=t$. Top: $n(\epsilon)$ obtained for several values of the interaction close to the phase transition, which takes place at $U_c=28.5$; inset: $n(k_x)$ for $k_y=k_z=0$ for the same parameters. Bottom: Comparison of the result for $n(\epsilon)$ close to the transition ($U=1.13 U_c$) obtained by different methods. The results obtained with the 3$^{rd}$ order strong coupling expansion in the hopping amplitude and the lattice QMC results are both from Ref.~\onlinecite{freericks}. }
\label{n_k}
\end{figure}

\subsubsection{Momentum distribution}

The momentum distribution function $n(k_x, k_y, k_z)$ of the normal  phase, calculated at $T=t$, is found to have an interesting behavior close to the transition to the superfluid phase. As shown in the upper panel and the inset of Fig.~\ref{n_k} the distribution $n(k_x)\equiv n(k_x,0,0)$ is strongly peaked at $k_x=0$ already in the normal phase.
In the B-DMFT the momentum dependence of the momentum distribution is expressed only through the non-interacting dispersion relation $\epsilon_{\bf k}$ (cf. Eq.~\eqref{g_k}). Therefore, $n(\epsilon)=n(\epsilon_{\bf k})$ implicitly determines the momentum distribution. The plots in Fig.~\ref{n_k} show $n(\epsilon)$ and $n(k_x)$ for different values of $U$ upon approaching the phase transition at constant density $\langle n \rangle = 1$.

The peak in the momentum distribution in the normal phase close to the phase transition was  noted previously by Kato \emph{et al.}~\cite{trivedi} within  QMC solution.
The lower panel in Fig.~\ref{n_k} shows a comparison between $n(\epsilon)$ obtained for the same parameters using different methods. As pointed out by Freericks \emph{et al.}~\cite{freericks} the increase in the occupation at $\epsilon=0$ is an effect which is only partially described by a strong-coupling expansion in the hopping amplitude. The B-DMFT does capture this enhancement, and our LCE results are in a very good agreement with the lattice QMC data of Ref.~\onlinecite{freericks}.

\begin{figure}
        \includegraphics[width=0.48\textwidth]{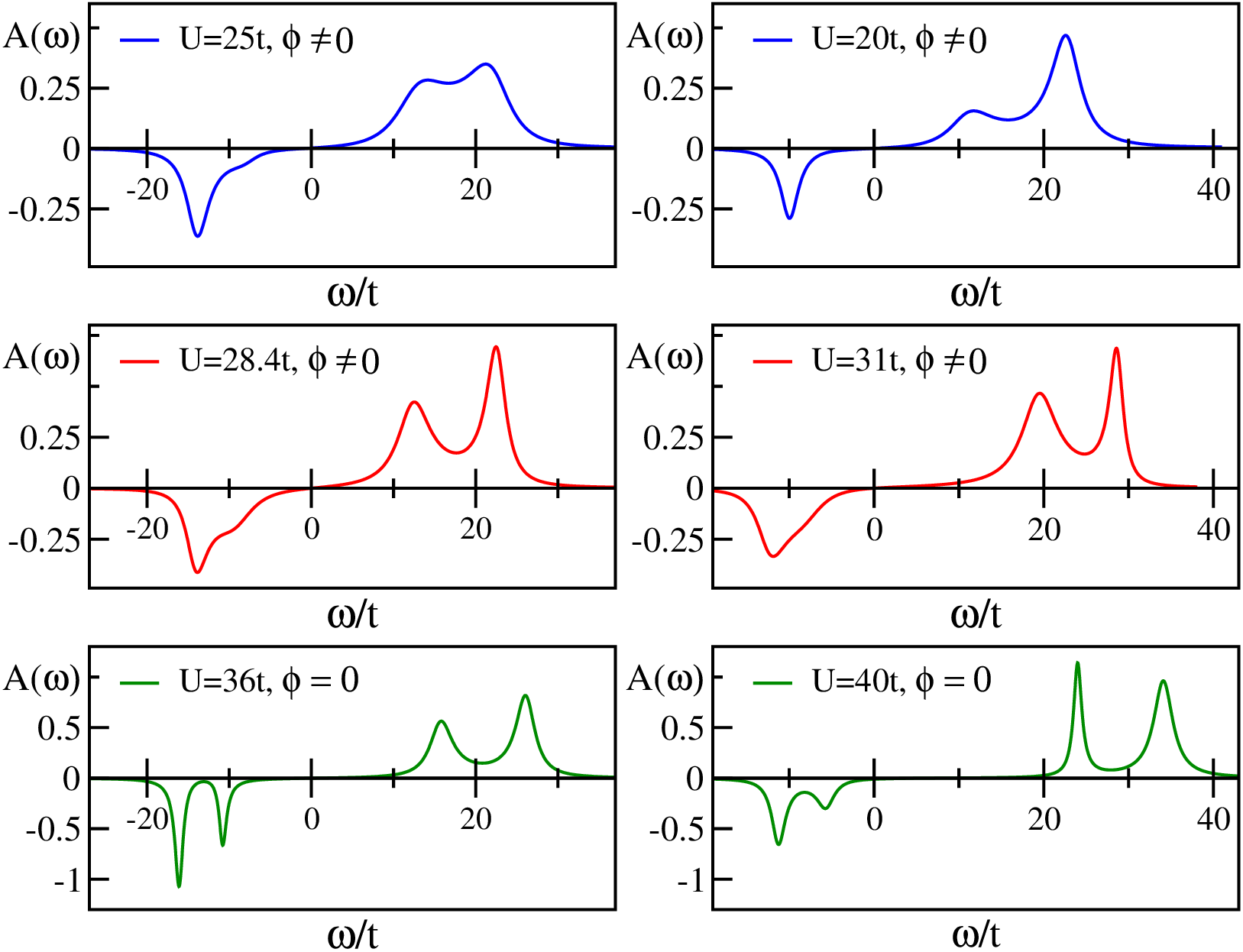}
\caption{Evolution of the spectral function (in arbitrary units) across the interaction driven phase transition at $T=2t$; left column: phase transition at the tip of the Mott lobe, $\langle n\rangle=1$; right column: phase transition away from the tip, $\mu=0.23 U$. In both columns the bottom plot is for the normal phase, whereas the two upper plots are for the superfluid phase. The energy scale is plotted relative to the chemical potential which is at $\omega=0$.\label{a_int}
}
\end{figure}

\subsubsection{Spectral functions}

\begin{figure}
        \includegraphics[width=0.29\textwidth]{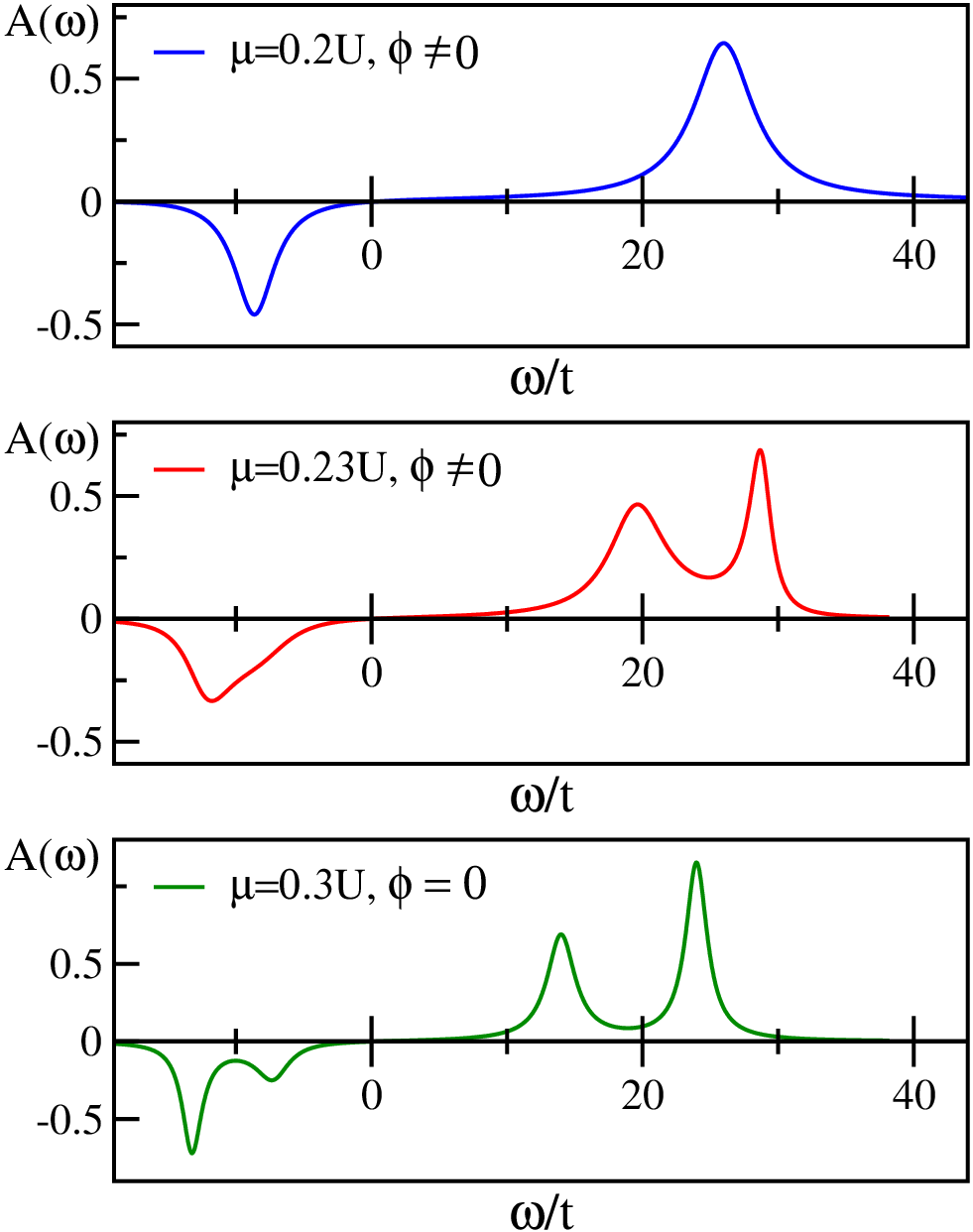}
\caption{Evolution of the spectral function (in arbitrary units) across the density driven phase transition at $U=31t$ and $T=2t$. The bottom plot is for the normal phase, whereas the two upper plots are for the superfluid phase. The  chemical potential is at $\omega=0$.\label{a_dens}}

\end{figure}

The B-DMFT approach with the LCE solver also allows one to investigate the behavior of the $k$-integrated spectral function $A(\omega)=-\frac{1}{\pi}\sum_{\bf k} Im G({\bf k}, \omega)$ across the phase transition from the superfluid to the Mott phase (Figs. ~\ref{a_int} and \ref{a_dens}).
 Since in our current implementation of the LCE the computations are performed on the imaginary time or imaginary frequency axes, spectral functions at real frequencies have to be calculated by analytic continuation.\cite{cm1} The spectral functions presented in Figs~\ref{a_int} and \ref{a_dens} were obtained by analytic continuation with Pad\'e approximants.
Calculations of bosonic spectral functions were also done with the functional renormalization group~\cite{kopietz,kopietz2} and in the variational cluster approach (VCA).\cite{knap,knap2}

Here we focus on three distinct cases: (i) the interaction driven phase transition at the tip of the Mott lobe, keeping the ratio $\mu/U$  constant; (ii) the interaction driven transition at the bottom of the lobe, also with  $\mu/U$ constant; and (iii) the density driven transition at constant interaction $U$.
Due to the approximation introduced by the analytic continuation one can draw only qualitative conclusions about the spectral density in the region close to the chemical potential (e.g., one cannot reliably estimate the size of the gap). Nevertheless the qualitative behavior and the spectral weight transfer is well illustrated and the difference between the three cases considered here is clearly visible. At the tip of the Mott lobe (case (i), left panel of Fig.~\ref{a_int}) an increase of the interaction leads to a symmetric shift of the spectral weight on both sides of the chemical potential. At the same time a Mott gap opens and two Hubbard bands are formed (see the bottom plot in the left panel of Fig.~\ref{a_int}). The shape of the bands vaguely resembles the non-interacting density of states $N_0(\epsilon)$ for the cubic lattice. Away from the tip (case (ii), right panel of Fig.~\ref{a_int}) the shift of the spectral weight is not symmetric with respect to the chemical 
 potential. The lower Hubbard band resides close to the chemical potential, whereas the upper Hubbard band is shifted to higher frequencies. A different behavior is observed in the density driven transition (case (iii), Fig.~\ref{a_dens}). Upon increasing the chemical potential at constant interaction, the spectral function is shifted as a whole to lower frequencies, simultaneously forming a gap.

\subsection{The Bethe lattice with $z=\infty$}

\begin{figure}
        \includegraphics[width=0.45\textwidth]{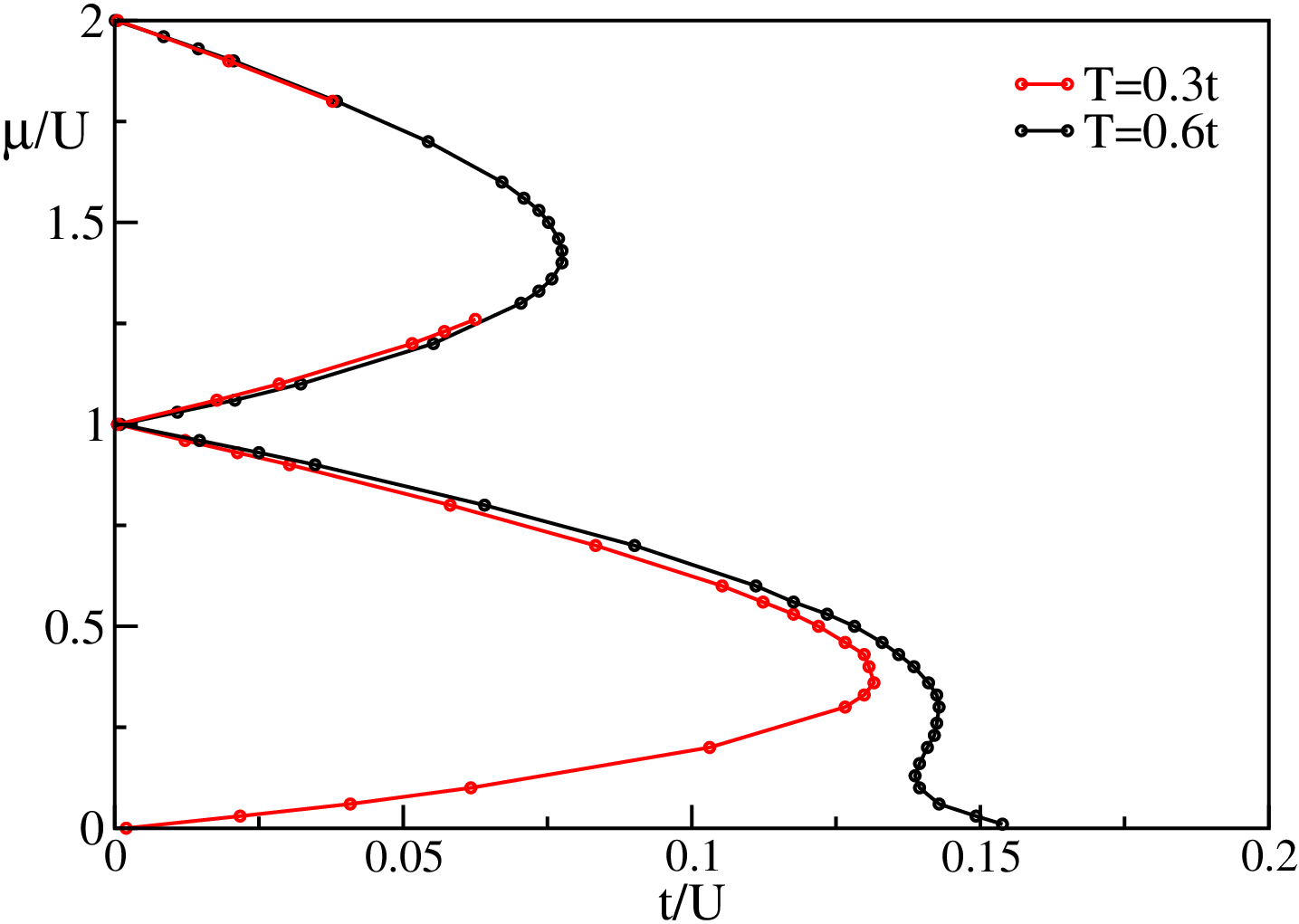}
\caption{Phase diagram $\mu/U$ \emph{vs.} $t/U$  for the $z=\infty$ Bethe lattice at two different temperatures. At $T=0.3t$ it was not possible to determine the phase boundary of the second lobe in the region $1.26U<\mu<1.8U$ (see the discussion in the text).}
\label{bethe_diagram}
\end{figure}

The phase diagram for the $z=\infty$ Bethe lattice is presented in Fig.~\ref{bethe_diagram}. At sufficiently high temperatures (e.g., $T=0.6t$ as in Fig.~\ref{bethe_diagram}) the LCE gives convergent results both for the superfluid and normal phases near the phase transition. However,
at temperatures below $0.4t$  we have not been able to find a convergent solution in the superfluid phase around the tip of the second Mott lobe. The iterations converge either to $\phi=0$ (normal phase), or to a solution with  $\phi\neq0$ but with a non-concave, and hence unphysical,\cite{cm3} $G_{11}(\tau)$. The results for $T=0.3t$ are shown in Fig.~\ref{bethe_diagram}, where the solution at $\mu/U$ around the first Mott lobe  converges both in the normal and the superfluid phase, thus making it possible to calculate the phase boundary. In the range $1.26U<\mu<1.8U$ a convergent solution was only obtained in the normal phase ($\phi=0$), i.e., it was not possible to determine the phase boundary of the second lobe completely.
As the temperature is lowered, the range of the chemical potentials for which we did not obtain a superfluid  solution increases. For temperatures below $0.3t$ we did not even obtain solutions with non-zero superfluid order parameter around the tip of the first Mott lobe. Upon further lowering the temperature, the region of convergence of the method in the superfluid phase is reduced to the values of $\mu$ near the edges of the lobes.
At this moment it is not clear whether the absence of a solution in the superfluid phase in the $z=\infty$ Bethe lattice for some chemical potentials at low temperatures is a consequence of the strong-coupling approximation to the B-DMFT, or the B-DMFT itself. This is an open question which needs to be answered in the future. Such problems did not occur for the other lattices investigated here.

\section{Summary}

 We developed an analytical approximation scheme to solve the B-DMFT equations for correlated lattice bosons in the strong-coupling limit. The solution makes use of a linked-cluster expansion in the hybridization function of normal bosons around the atomic limit. Explicit results were obtained for the Bose-Hubbard model on the cubic lattice and the Bethe lattice with connectivity $z=6$ and $z=\infty$, respectively. Remarkably good agreement with numerical solutions of the
B-DMFT equations obtained with exact diagonalization~\cite{hofstetter}, continuous-time
quantum Monte Carlo~\cite{werner2}, and direct lattice QMC calculations~\cite{qmc}  was found.
This agreement demonstrates that the strong-coupling solution derived here provides a correct description of the physics of correlated bosons. The method is computationally inexpensive and, with a good choice of the initial guess of the parameters, usually leads to a fast convergence of the iteration of the self-consistency equations. The Bethe lattice with infinite connectivity is an exception which still requires further investigation.

We also employed the linked-cluster expansion to calculate the momentum distribution function of normal bosons close to the phase transition as well as the bosonic spectral function in the normal and superfluid phase.

The approximation scheme presented in this paper can, in principle, be systematically improved by the inclusion of higher order terms. However, the non-interacting limit can only be reached if terms up to infinite order are included, e.g., by an appropriate resummation. This has been achieved for fermions by the non-crossing approximation (NCA).\cite{nca} The fundamental problem of the NCA, namely its failure to describe the low temperature Fermi liquid regime adequately owing to the existence of a characteristic coherence scale (the Kondo temperature), may be absent in the case of bosons where such a coherence scale does not exist. For that reason it should be clarified whether it is possible to construct a renormalized expansion for correlated bosons which is applicable for all temperatures and interaction strengths.

\begin{acknowledgments}
This research was done when Anna Kauch worked at the University of Augsburg. Partial support by the Deutsche Forschungsgemeinschaft through TRR 80 is gratefully acknowledged. Krzysztof Byczuk acknowledges support by the grant No. N~N202 103138 of the Polish Ministry of Science and Education.
\end{acknowledgments}


\begin{thebibliography}{41}
\bibitem [{cm2()}]{cm2}%
  \BibitemOpen
  \href@noop {} {}\bibinfo {note} {Results presented here were usually computed
  with the maximal occupation number equal to 5, which introduces an error much
  below the resolution of the rest of the calculations.}\BibitemShut {Stop}%
\bibitem [{cm1()}]{cm1}%
  \BibitemOpen
  \href@noop {} {}\bibinfo {note} {Our experience is that the convergence is
  much faster in the imaginary time implementation, so we have dropped the real
  frequency approach for the moment.}\BibitemShut {Stop}%
\bibitem [{cm3()}]{cm3}%
  \BibitemOpen
  \href@noop {} {}\bibinfo {note} {It follows from the definition of the
  diagonal part of the bosonic Green function in imaginary time that it has to
  be concave.}\BibitemShut {Stop}%
\end{thebibliography}
\end{document}